\documentclass[twocolumn,aps,prl,ltxgrid]{revtex4}

\usepackage{epsfig}
\usepackage{graphicx}
\usepackage{amsbsy,amssymb,amsmath,bm,ulem}
\usepackage[usenames]{color}

\newcommand{\f}[1]{Fig.~\ref{#1}}

\newcommand{\eqs}[2]{Eqs.~(\ref{#1}) and~(\ref{#2})}

\newcommand{\Ht}{H^{\text{th}}}
\newcommand{\Htt}{H_1^{\text{th}}}
\newcommand{\Httt}{H_2^{\text{th}}}

\begin{document}

\title{Reentrant stability of superconducting films}

\author{V. V. Yurchenko, D. V. Shantsev and T. H. Johansen}
\affiliation{Department of Physics and Center for Materials
Science and Nanotechnology,    University of Oslo, P. O. Box 1048
Blindern, 0316 Oslo, Norway}

\author{M. R. Nevala, I. J. Maasilta}
\affiliation{Nanoscience Center, Department of Physics,
University of Jyvaskyla, P. O. Box 35, FIN 40014, Finland}

\author{K. Senapati, R. C. Budhani}
\affiliation{Department of Physics, Indian Institute
of Technology Kanpur, Kanpur 208016, India}

\begin{abstract}
\vspace{2mm} \noindent We propose a mechanism responsible for the
abrupt vanishing of the dendritic flux instability found in many
superconducting films when an increasing magnetic field is applied.
The onset of flux avalanches and the subsequent reentrance of
stability in NbN films was investigated using magneto-optical
imaging, and the threshold fields were measured as functions of
critical current density, $j_c$. The results are explained with
excellent quantitative agreement by a thermomagnetic model published
recently,
 \prb \textbf{73}, 014512 (2006), showing that the reentrant
stability is a direct consequence of a monotonously decreasing $j_c$
versus field.

\end{abstract}

\maketitle

\vspace{8mm}

In most superconductors a slow increase of external magnetic
field is accompanied by a gradual penetration of magnetic vortices,
which make their way into the sample through a random energy
landscape created by structural irregularities. This results in
formation of a critical state characterized by a critical gradient
of the vortex density corresponding to the maximum lossless current
in the superconductor.\cite{Bean} The critical state, however, is
metastable and can be destroyed by  flux jumps where large-scale
redistribution of the vortices suddenly takes place. Such dramatic
events, observed experimentally as abrupt
drops in the magnetization,
are due to a thermo-magnetic instability where the local dissipation
associated with vortex motion reduces the pinning, which in turn
facilitate further motion.\cite{Mints81} With this positive feedback
a small perturbation can quickly evolve into a flux avalanche of
sample spanning dimensions.

In recent years, space-resolved measurements, in particular using
magneto-optical imaging (MOI), have revealed the detailed morphology
of flux avalanches. Most work was carried out on superconducting
thin
films of MgB$_2$,\cite{Johansen02,Barkov03,ye04,Albrecht,Laviano} Nb,\cite{Duran95,welling04} Pb,\cite{menghini}
Nb$_3$Sn,\cite{Rudnev03} NbN,\cite{Rudnev05}
YBa$_2$Cu$_3$O$_x$,\cite{Leiderer93,biehler} and
YNi$_2$B$_2$C,\cite{Wimbush}  where the images show that a typical
flux  avalanche in superconducting films has a branched dendritic
structure with 50-100~micron wide branches. While each avalanche
event has only local impact, they occur very frequently, i.e., with
very small applied field intervals, and therefore manifest in
magnetization versus field curves as a strong noisy
component.\cite{Johansen02,Wimbush,ye04,Rudnev05,r3,r5}

These magnetization curves display also two other generic features,
namely the existence of a threshold field for the onset of avalanche
activity, typically at a few millitesla, and an upper threshold
field above which the superconductor regains full stability.
 In a recent Letter\cite{Denisov-prl} we reported a detailed study of
the lower threshold, and explained key features of the instability
onset. The important question why superconducting films suddenly
become stable again at high fields remains open, and is the focus of
the present work. We report here experiments carried out on films of
superconducting NbN, and discuss the results within the theoretical
framework of a recently proposed thermo-magnetic
model\cite{Denisov06,Aranson}. The reentrant stability is explained
by accounting for the field dependence of the critical current
density.



Superconducting films of NbN were prepared on single crystal (100) MgO
substrates by pulsed laser ablation of a high purity Nb target in a controlled ultra high
purity (99.9996) N$_2$ environment. A KrF excimer laser (248 nm)
operated at 20 Hz was used for ablation with pulse energy density of
5~J/cm$^2$ on the target surface. Detailed procedures and structural characterization of the samples, which are
superconducting below $T_c = 15$~K, is found in
Ref.~\onlinecite{Senapati}. The films have a thickness of
$d=280$~nm, and were lithographically patterned into a rectangular
shape 2.4 mm wide and 4.8 mm long  using reactive ion etching in a
CHF$_3$ + O$_2$ plasma.

Flux penetration into the samples was investigated using a
MOI method based on the Faraday effect in an in-plane magnetized
bismuth-substituted ferrite garnet sensor film, that was placed on top of the NbN film. The sample was mounted on the cold finger
of a continuous He-flow cryostat (Microstat, Oxford), which in turn was placed under the objective
lens of a reflected light polarizing microscope (Leica). A variable magnetic field was applied
perpendicular to the sample using a pair of resistive coils. The microscope was equipped with a mercury lamp,
and by using crossed polarizers the resulting image brightness represents the local magnitude of the perpendicular
flux density.

%
\begin{figure}[t]
  \epsfig{file=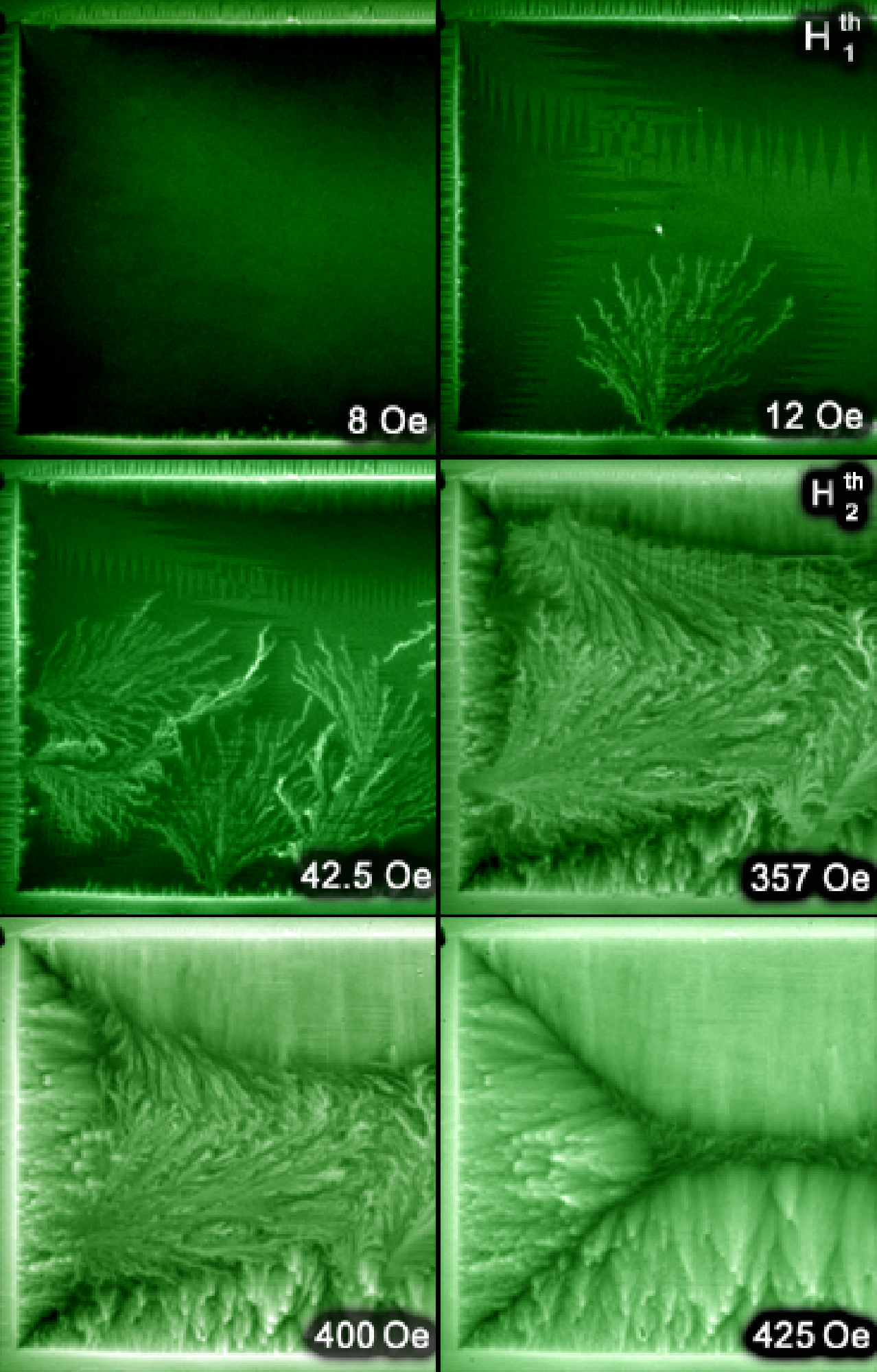, width=8cm}
  \caption{Magneto-optical images of a NbN film (only left half of the sample area is shown) at 4~K in
  increasing applied field. The left edge of the sample, seen as a vertical line of enhanced brightness, is 2.4 mm
long. The images were contrast enhanced individually.}
  \label{f_mo}
\end{figure}

Shown in \f{f_mo} is a set of images recorded during a slow ramping
up of the applied magnetic field, $H$, after the sample was
initially zero-field cooled to $T = 4$~K. In image (a), taken at $H = 8$~Oe, there is only a shallow penetration, with a flux
front that moved gradually inwards as the field increased. This
smooth mode of penetration ended abruptly when
reaching $H = 12$~Oe, where a flux avalanche suddenly occurred. The avalanche area is seen
in \f{f_mo} taken at 12 Oe as the branching structure rooted on the lower sample edge. As the
applied field increased further, more and more avalanches took place, creating a complex pattern of flux dendrites
covering most of the sample area, see \f{f_mo} at $H =42.5$~Oe. Then, as the
field reached $H =357$~Oe, the avalanche activity stopped entirely although the field continued to increase.
The advancing flux front now erased the previously formed
dendritic structures, see  \f{f_mo} at $H =400$~Oe and $425$~Oe, and  the critical state mode of flux penetration was restored.
The bright fan-like features in the lower part of the images are due to tiny film defects, and are not related to the flux instability. 


This experiment gives direct evidence that there exists an upper
threshold field for the dendritic instability, in agreement with
earlier measurements on various superconducting
films.\cite{Johansen02,r5,Wimbush,Rudnev05} Before presenting
more results, we give first a qualitative argument for why reentrant
stability can follow from the thermo-magnetic model proposed in
Ref.~\onlinecite{Denisov06}. The theory predicts that a
 superconducting thin strip of half-width $w$ 
becomes unstable when the flux penetration depth exceeds
\begin{equation}
\ell^* =\frac{\pi}{2} \sqrt{\frac{\kappa T^*}{j_c E}} \left( 1 -
\sqrt{\frac{2h_0 T^*}{nd j_c E}}\right)^{-1}\, ,
\label{l}
\end{equation}
provided that $\ell^* < w$. Here, $j_c$ is the critical current
density, $T^*\equiv -(\partial \ln j_c/\partial T)^{-1}$, $E$ is
the electric field, $\kappa$~is the thermal conductivity, and
$h_0$ is the coefficient of heat transfer from the superconducting
film to the substrate. The parameter $n$ characterizes the
nonlinearity of the current-voltage curve of the superconductor,
$n=\partial \ln E/\partial \ln j \gg 1$. The instability onset
field can be obtained from $\ell^*$ using the Bean model relation
between the penetration depth and applied field,\cite{BrIn}
\begin{equation}
\Ht =  \frac{j_c d}{\pi} \;  {\rm arccosh}
\left(\frac{w}{w-\ell^*} \right) \ .\label{lH}
\end{equation}
\begin{figure}[t]
  \epsfig{file=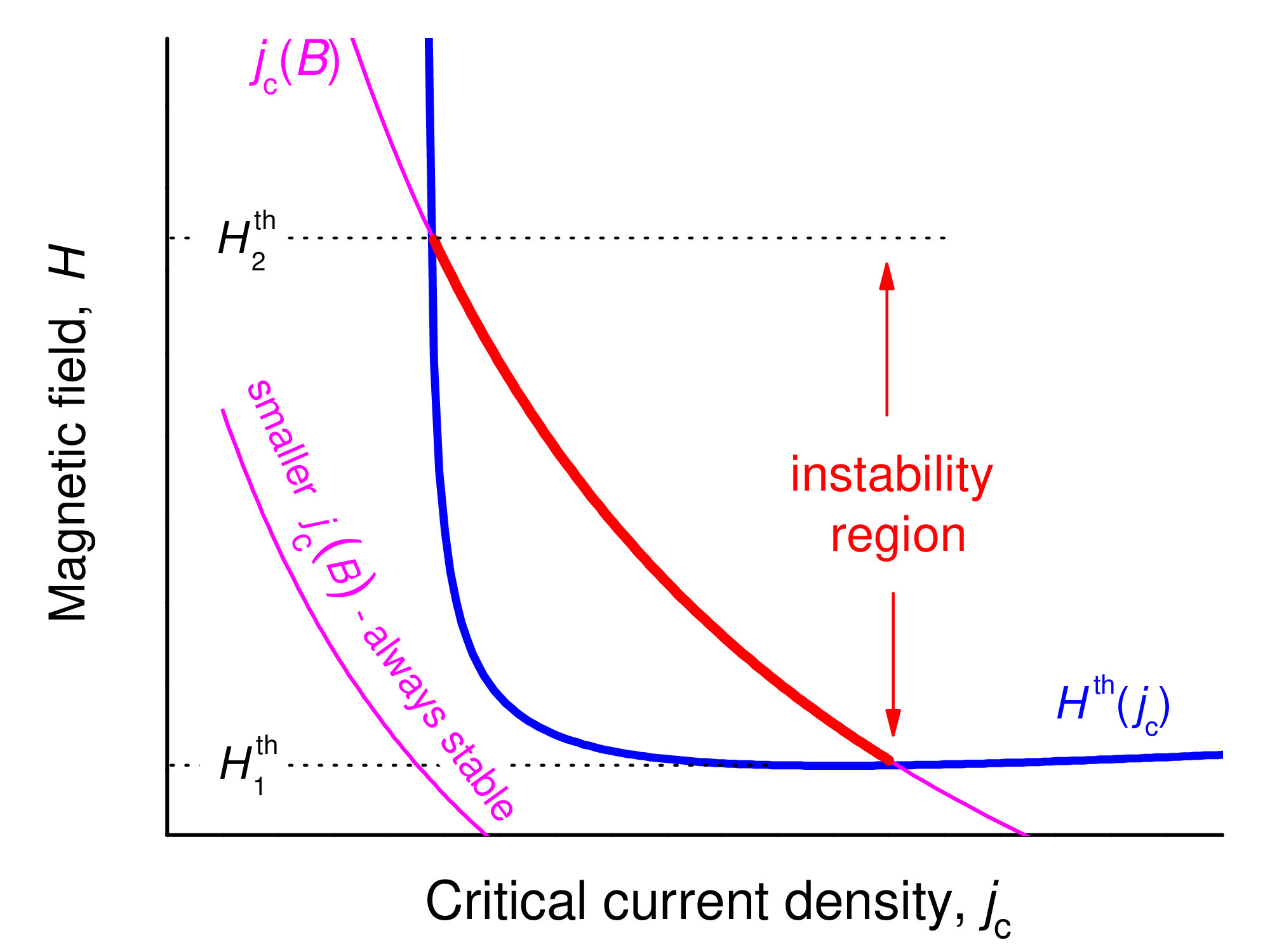, width=8cm}
  \caption{Schematic plot illustrating the existence of two
  threshold fields for the dendritic
  instability. The two main curves represent the
  threshold field $\Ht$ versus $j_c$ according to
\eqs{l}{lH}, and a typical monotonous field dependence of the
critical current density. The intersection of the two curves defines
the two thresholds $\Htt$ and $\Httt$ for the onset and
vanishing of the instability.}
  \label{f_idea}
\end{figure}
It follows from \eqs{l}{lH} that the threshold field depends
strongly on the critical current density, more specifically
according to the graph marked $\Ht(j_c)$ in \f{f_idea}. For
intermediate $j_c$ the threshold field is nearly constant, and
increases slowly as $j_c$ becomes larger giving eventually a
sublinear asymptotic dependence, $\Ht \sim j_c^{3/4}$. More
importantly, when $j_c$ decreases the threshold field will diverge
at some finite $j_c$, corresponding to $\ell^*(j_c)$ approaching
$w$. Taking now into account that the critical current density most
commonly decreases with the field, typically as the $j_c(B)$ curve
in \f{f_idea}, it follows that the field range with unstable
behavior can indeed have both a lower and upper limit, $\Htt$ and
$\Httt$, as indicated in the figure. Evidently, it is essential in
this picture how fast $j_c$ is decreasing with the field. E.g., in
the Bean model (constant $j_c$), or with a $j_c$ having only weak
field dependence, the conditions for having an upper threshold will
never be met. Note also that if $j_c$ is sufficiently small, e.g.,
because of a larger temperature, the thermo-magnetic avalanches will
not occur at any magnetic field.


To obtain quantitative support for this  explanation we performed
MOI of flux penetration after first cooling the sample to 4~K in
various constant magnetic fields, $H_{\rm fc}$. This allowed
measurement of the field dependence of both $j_c$ and the two
instability threshold fields. Cooling fields up to 300~Oe were
applied, and MOI did not detect any
 contrast, which implies that the full flux density $H_{\rm
fc}$ was frozen into the sample. To measure $j_c$ a small additional
field was subsequently applied, creating a critical state type of
penetration of the new flux. 
\endnote{To avoid flux dendrite formation in these experiments
the superconductor was placed in close contact with the metallic
mirror of the garnet sensor film, which is known to suppress the
instability.\protect\cite{Baz02,Korea} } From the depth of the penetration
front in the middle section of the rectangular sample the $j_c$ was
determined,\protect\cite{BrIn} and \f{f_jc} shows its dependence on the
cooling field. A nearly exponential decay was found, with $j_c$
decreasing from $7.6 \cdot 10^{10} $~Am$^{-2}$ at zero field to
almost one half at $H_{\rm fc} = 300$ Oe. This agrees well with
the results obtained earlier from ac susceptibility measurements
\cite{Senapati} on the same type of films.

\begin{figure}[t]
  \epsfig{file=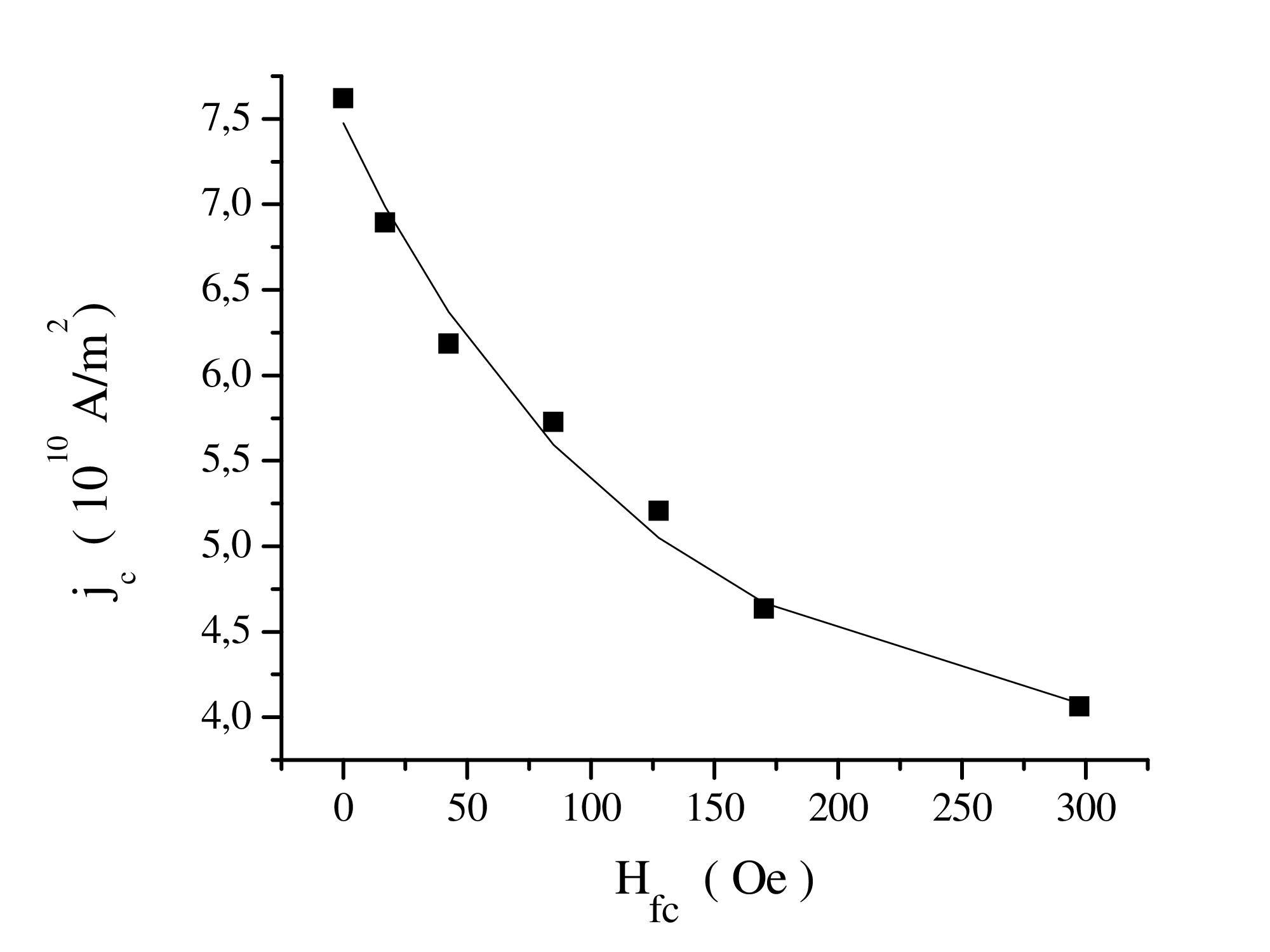, width=8cm}
  \caption{Field dependence of the critical current
  density at $T = 4$~K measured using MOI.}
  \label{f_jc}
\end{figure}

The threshold fields were determined by slowly ramping up the
additional field after an initial field-cooling. The values of the
total applied field when the first and last flux dendrite appeared
were recorded for various $H_{\rm fc}$, giving the result shown in
\f{f_hth}. The onset field increases almost linearly, and with some
upward curvature, whereas the upper threshold field remains
essentially constant until the two thresholds eventually merge into
one when $H_{\rm fc} = 300$ Oe. For larger cooling fields,
i.e., for smaller $j_c$'s, the instability was found to be fully
suppressed.

\begin{figure}[t]
  \epsfig{file=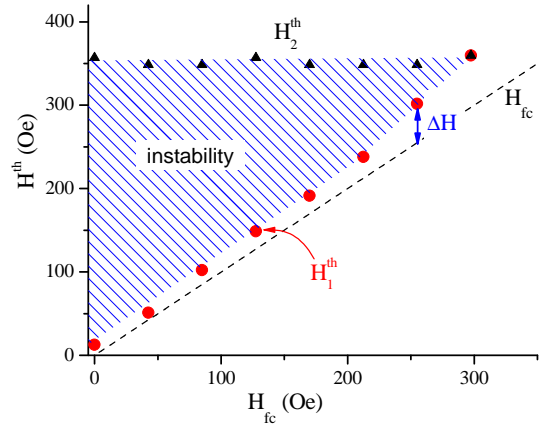, width=8cm}
  \caption{The lower and upper threshold fields $\Htt$ (disks) and $\Httt$ (triangles) measured at different
   frozen-in fields after field-cooling to 4 K.
  }
  \label{f_hth}
\end{figure}

To demonstrate that these data are in excellent agreement with the
scenario depicted in \f{f_idea}, we note first that the applied
field which induces shielding currents in the sample, and triggers
the first avalanche, is $\Delta H = \Htt - H_{\rm fc}$. Thus, in the
presence of a cooling field, the curve representing the $\Ht(j_c)$
in \f{f_idea} should be shifted upwards by the amount $H_{\rm fc}$,
while the $j_c(B)$ curve remains as is. Then, as $H_{\rm fc}$
becomes larger the distance between the two intersection points of
the curves becomes gradually smaller, and at some cooling field the
two thresholds become one, just like in the experiments. To show the
agreement quantitatively, the data for the lower threshold field are
replotted in \f{f_fit} as $\Delta H$ versus $j_c$. The full curve is
a theoretical curve obtained from \eqs{l}{lH}, using two fitting
parameters, $\kappa T^*/E=343.75$ W/V and $h_0T^*/nE=687.5$ kW/V (which can mean,
e.g., $\kappa = 5$ W/Km, $h_0 = 110$ kW/Km, $n=11$, $E= 0.16$ V/m, $T^*=11$ K ).

\begin{figure}[t]
  \epsfig{file=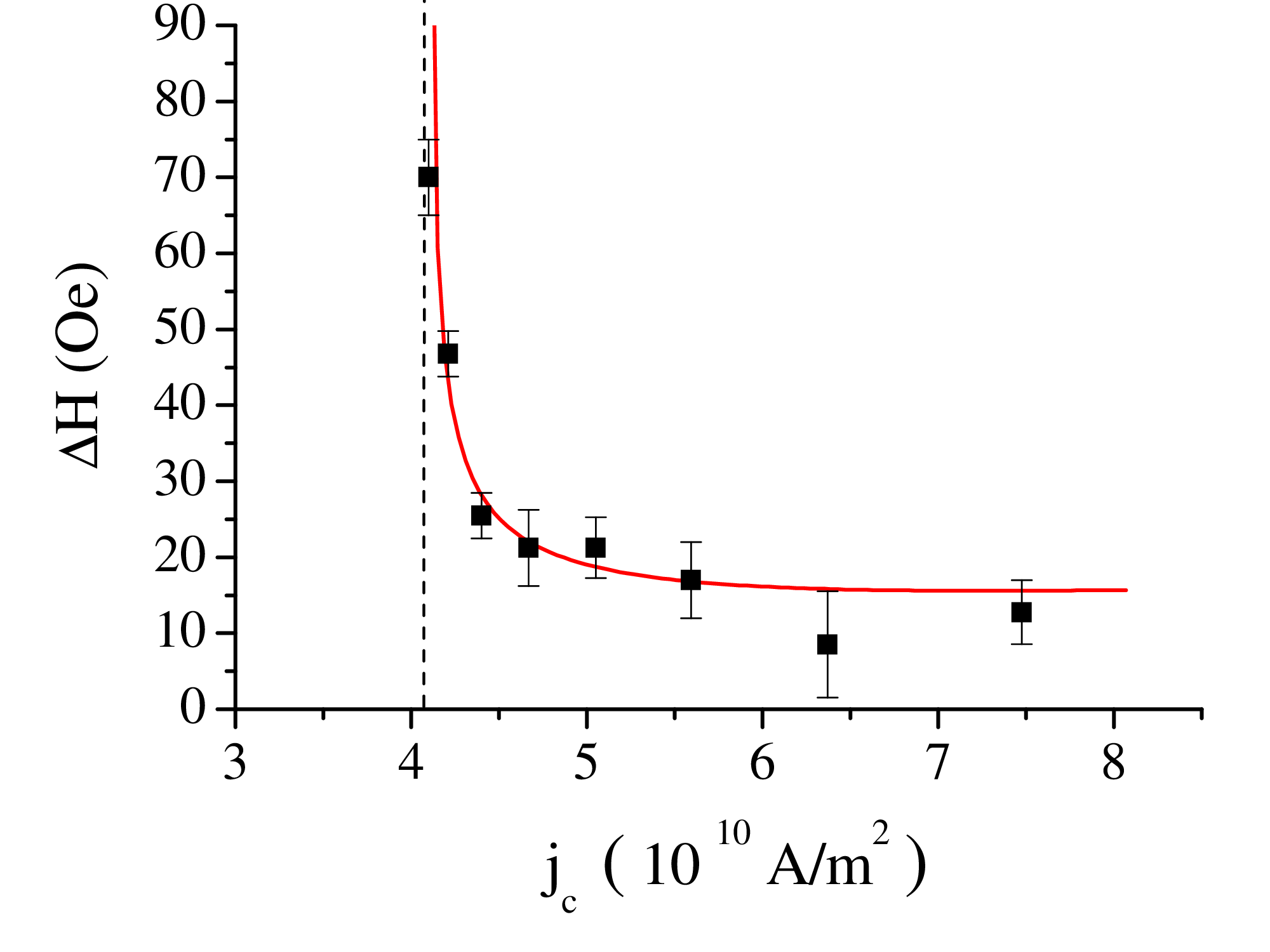, width=8cm}
  \caption{Instability onset field $\Delta H$
  as a function of the critical current at 4~K. The plot shows
  experimental data (squares) and a theoretical prediction
  (full line) using fitting parameters given in the text.
   This plot made use also of the experimental data in \f{f_jc},
   to convert from $H_{\rm fc}$ to $j_c$.
The error bars indicate the spread in the data obtained by repeating
the experiments several times.
  }
  \label{f_fit}
\end{figure}

In contrast to $\Htt$, the upper threshold field $\Httt$ was found
to be essentially independent of $H_{\rm fc}$. Even this behavior is
fully consistent with our explanation illustrated by \f{f_idea}.
Near the intersection point determining $\Httt$, the diverging
$\Ht(j_c)$ curve has a very steep slope. Hence, a vertical shift of
$\Ht(j_c)$ accounting for a non-zero cooling field $H_{\rm fc}$
hardly changes the value of $\Httt$. The robustness of the upper
threshold field by being independent of the initial conditions, is
actually a remarkable and unexpected result. Indeed, the flux
distribution after cooling the sample in $H_{\rm fc}$ and then
ramping the field up to $\Httt$ dramatically depends on the value of
$H_{\rm fc}$. In the zero-field cooled experiment shown in \f{f_mo}
one finds dozens of overlapping dendrites, while for $H_{\rm
fc}=300$~Oe the flux distribution remains of a critical-state type
until a single dendritic structure is formed at $\Httt$. Note that
the exact dendritic pattern is every time different even if one
repeats the experiment under the same conditions. Thus, despite the
wide variety of flux distributions, the dendritic activity in all
these cases vanishes at the {\em same} threshold field $\Httt
\approx 350$~Oe. Note also that the $\Ht(j_c)$ curve in \f{f_idea}
was calculated\cite{Denisov06} assuming  an initial critical-state
flux distribution, and therefore the observed invariance of $\Httt$
suggests that the theoretical prediction is valid for a much broader
set of initial magnetic states.

The present thermomagnetic scenario, and in particular the
$\Ht(j_c)$ dependence shown in \f{f_idea}, has very recently
received even further support from MOI studies of MgB$_2$ films
grown on vicinal substrates giving a slight anisotropy in the
critical current.\cite{Albrecht07} The observed dramatic anisotropy
in the direction of dendritic avalanches was fully explained by the
very steep slope of the $\Ht(j_c)$ curve.

In conclusion, the dendritic flux instability, which is a serious threat to superconducting 
devices based on thin films, was investigated to understand mechanisms that allow the films to recover
stability. We explained reentrant stability in the framework of a thermo-magnetic model, \cite{Denisov06} 
and demonstrated by a quantitative agreement of the theory with experimental results, that it stems from the field dependence of the critical current.

This work was supported by the Research Council of Norway, Grant
No. 158518/431 (NANOMAT). Research at IIT Kanpur has been
supported by a grant from the Council of Scientific and Industrial
Research (CSIR) Government of India. The authors acknowledge fruitful discussions with D.G. Gheorghe and R.J. Wijngaarden.\\

\bibliographystyle{revtex}

\begin{thebibliography}{99}

\bibitem{Bean} C. P. Bean, Rev. Mod. Phys. {\bf 36}, 31 (1964).


\bibitem{Mints81} R. G. Mints and A. L. Rakhmanov, Rev. Mod. Phys.
\textbf{53}, 551 (1981).


\bibitem{Johansen02} T. H. Johansen {\it et al.},
Europhys. Lett.\textbf{59}, 599 (2002).

\bibitem{Barkov03} F. L. Barkov, 
D. V. Shantsev, T. H. Johansen, P. E. Goa, W. N. Kang, H. J. Kim, E.
M. Choi, S. I. Lee, Phys. Rev. B \textbf{67}, 064513 (2003).


\bibitem{ye04} Z. X. Ye 
Q. Li, Y. F. Hu, A. V. Pogrebnyakov, Y. Cui, X. X. Xi, J. M.
Redwing, and Q. Li Appl. Phys. Lett. {\bf 85}, 5284 (2004).

\bibitem{Albrecht} J. Albrecht {\it et al.},
Appl. Phys. Lett. \textbf{87}, 182501 (2005).

\bibitem{Laviano} F. Laviano {\it et al.}, 
in Magneto-Optical Imaging, edited by T. H. Johansen and D. Shantsev
(Kluwer Academic, 2004), p. 237.


\bibitem{Duran95} C. A. Duran, P. L. Gammel, R. E. Miller, D. J.
Bishop, Phys. Rev. B \textbf{52}, 75 (1995).


\bibitem{welling04} M. S. Welling, R. J. Westerwaal, W. Lohstroh,
R. J. Wijngaarden, Physica C {\bf 411}, 11 (2004).

\bibitem{menghini} M. Menghini, 
R. J. Wijngaarden, A. V. Silhanek, S. Raedts and V. V. Moshchalkov,
Phys. Rev. B {\bf 71}, 104506 (2005).


\bibitem{Rudnev03} I. A. Rudnev, 
S. V. Antonenko, D. V. Shantsev, T.~H.~Johansen, A. E. Primenko,
Cryogenics \textbf{43}, 663 (2003).

\bibitem{Rudnev05} I. A. Rudnev, D. V. Shantsev, T. H. Johansen, A. E.
Primenko, Appl. Phys. Lett. {\bf 87}, 042502 (2005)

\bibitem{Leiderer93} P. Leiderer, 
J. Boneberg, P. Br¨ull, V. Bujok, S. Herminghaus, Phys. Rev. Lett.
\textbf{71}, 2646 (1993).

\bibitem{biehler} B. Biehler, 
B.-U. Runge, P. Leiderer, and R. G. Mints, Phys. Rev. B \textbf{72},
024532 (2005).

\bibitem{Wimbush} S. C. Wimbush, B. Holzapfel,
and Ch. Jooss, J. Appl. Phys. {\bf 96}, 3589 (2004).

\bibitem{r5}
Zhao Z. W. {\it et al.}, \prb \textbf{65}, 064512 (2002).


\bibitem{r3}
Jin S., Mavoori  H.,  Bower C. \and van Dover R. B. {Nature} {\bf
411}, {563} ({2001}).

\bibitem{Denisov-prl}  D. V. Denisov {\it et al.}, 
Phys. Rev. Lett. {\bf 97}, 077002 (2006).

\bibitem{Aranson} I. S. Aranson, 
A. Gurevich, M. S. Welling, R. J. Wijngaarden, V. K. Vlasko-Vlasov,
V. M. Vinokur, U. Welp, \prl \textbf{94}, 037002 (2005)

\bibitem{Denisov06} D. V. Denisov, A. L. Rakhmanov, D. V. Shantsev,
Y. M. Galperin, T. H. Johansen, \prb \textbf{73}, 014512 (2006).


\bibitem{Senapati}  K. Senapati, N. K. Pandey, R. Nagar, and R. C. Budhani, Phys. Rev. B 74, 104514 (2006).



\bibitem{jooss} S. C. Wimbush, B. Holzapfel, Ch. Jooss, J. Appl.
Phys. {\bf 96}, 3589 (2004).


\bibitem{Baz02} M. Baziljevich, A. V. Bobyl, D. V. Shantsev, E. Altshuler, T. H. Johansen, and S. I. Lee,
Physica C {\bf 369}, 93 (2002).

\bibitem{Korea} Eun-Mi Choi {\it et al.},
Appl. Phys. Lett. \textbf{87}, 152501 (2005).


\bibitem{Albrecht07} J. Albrecht et al., Phys. Rev. Lett. in press,
cond-mat/0607063.



\bibitem{BrIn}E. H. Brandt, and M. Indenbom, Phys. Rev. B {\bf 48},12893
(1993).

\end{thebibliography}

\end{document}